\documentclass[mathleft
]{an}
\usepackage{graphicx}
\usepackage{times}
\usepackage{natbib}
\bibpunct{(}{)}{;}{a}{}{}
\begin{document}

\Pagespan{789}{}
\Yearpublication{2006}%
\Yearsubmission{2005}%
\Month{11}%
\Volume{999}%
\Issue{88}%

\def\aj{AJ}%
\def\actaa{Acta Astron.}%
\def\araa{ARA\&A}%
\def\apj{ApJ}%
\def\apjl{ApJ}%
\def\apjs{ApJS}%
\def\ao{Appl.~Opt.}%
\def\apss{Ap\&SS}%
\def\aap{A\&A}%
\def\aapr{A\&A~Rev.}%
\def\aaps{A\&AS}%
\def\azh{AZh}%
\def\baas{BAAS}%
\def\bac{Bull. astr. Inst. Czechosl.}%
\def\caa{Chinese Astron. Astrophys.}%
\def\cjaa{Chinese J. Astron. Astrophys.}%
\def\icarus{Icarus}%
\def\jcap{J. Cosmology Astropart. Phys.}%
\def\jrasc{JRASC}%
\def\mnras{MNRAS}%
\def\memras{MmRAS}%
\def\na{New A}%
\def\nar{New A Rev.}%
\def\pasa{PASA}%
\def\pra{Phys.~Rev.~A}%
\def\prb{Phys.~Rev.~B}%
\def\prc{Phys.~Rev.~C}%
\def\prd{Phys.~Rev.~D}%
\def\pre{Phys.~Rev.~E}%
\def\prl{Phys.~Rev.~Lett.}%
\def\pasp{PASP}%
\def\pasj{PASJ}%
\def\qjras{QJRAS}%
\def\rmxaa{Rev. Mexicana Astron. Astrofis.}%
\def\skytel{S\&T}%
\def\solphys{Sol.~Phys.}%
\def\sovast{Soviet~Ast.}%
\def\ssr{Space~Sci.~Rev.}%
\def\zap{ZAp}%
\def\nat{Nature}%
\def\iaucirc{IAU~Circ.}%
\def\aplett{Astrophys.~Lett.}%
\def\apspr{Astrophys.~Space~Phys.~Res.}%
\def\bain{Bull.~Astron.~Inst.~Netherlands}%
\def\fcp{Fund.~Cosmic~Phys.}%
\def\gca{Geochim.~Cosmochim.~Acta}%
\def\grl{Geophys.~Res.~Lett.}%
\def\jcp{J.~Chem.~Phys.}%
\def\jgr{J.~Geophys.~Res.}%
\def\jqsrt{J.~Quant.~Spec.~Radiat.~Transf.}%
\def\memsai{Mem.~Soc.~Astron.~Italiana}%
\def\nphysa{Nucl.~Phys.~A}%
\def\physrep{Phys.~Rep.}%
\def\physscr{Phys.~Scr}%
\def\planss{Planet.~Space~Sci.}%
\def\procspie{Proc.~SPIE}%
\let\astap=\aap
\let\apjlett=\apjl
\let\apjsupp=\apjs
\let\applopt=\ao

\def\rxj{RX\,J0720.4-3125 }
\def\mmas{\mathrm{mas}}
\newcommand{\xmm}{XMM-Newton}
\title{New photometry and astrometry of the isolated neutron star \rxj using recent VLT/FORS observations\thanks{Based on observations made with ESO Telescopes at the Paranal Observatory under programme ID 080.D-0135(A). Based on observations with XMM-Newton, an ESA Science Mission with 
instruments and contributions directly funded by ESA Member states and the USA (NASA).}}

\author{Thomas Eisenbeiss\inst{1}\fnmsep\thanks{Corresponding author: \email{eisen@astro.uni-jena.de}\newline}
\and{Christian Ginski}\inst{1}
\and{Markus M. Hohle}\inst{1,2}
\and Valeri V. Hambaryan\inst{1}
\and Ralph Neuh\"auser\inst{1}
\and Tobias O. B. Schmidt\inst{1}
}
\titlerunning{Vmag and position of \rxj}
\authorrunning{T. Eisenbeiss et al. }
\institute{Astrophysikalisches Institut und Universit\"ats-Sternwarte Jena, Friedrich-Schiller-Universit\"at Jena, Schillerg\"asschen 2-3, 07745 Jena \and Max Planck Institut f\"ur extraterrestrische Physik Garching, Giessenbachstrasse, 85738 Garching}
\received{}
\accepted{}
\publonline{}

\keywords{Stars: neutron - Dense matter - Stars: imaging - Techniques: image processing - { Stars: individual}}

\abstract{%
Since the first optical detection of \rxj various observations 
{ have been} 
performed to 
{ determine} 
astrometric and photometric data. We 
{ present}   
the first detection of the isolated neutron star in the V Bessel filter to study the spectral energy distribution and { derive} a new astrometric position.
At ESO Paranal  we obtained very deep images with FORS 1 (three hours exposure time) of \rxj in V Bessel filter in January 2008. We derive the visual magnitude by standard star aperture photometry.
Using sophisticated resampling software we correct the images for field distortions. Then we derive an updated position and proper motion value by comparing its position with FORS 1 observations of December 2000.
We { calculate} a visual magnitude of $V=26.81\pm0.09$\,mag, which is seven times in excess 
{ of} 
what is expected from X-ray data, but consistent with 
{ the extant} 
U, B and R data.
{ Over} 
about 
{ a} 
seven
year  
epoch difference we measured a proper motion of $\mu=105.1\pm7.4$\,mas\,yr$^{-1}$ towards $\theta=296.951^{\circ}\pm0.0063^{\circ}$ (NW)
,  
 consistent with previous data.}

\maketitle

\section{Introduction}
In 
deep optical follow-up observations of bright X-ray sour\-ces from the ROSAT mission \citep{1999A&A...349..389V} seven thermal emitting { isolated} neutron stars (INS), { exhibiting no radio emission}, 
{ have been}  
discovered so far (see \citealp{2004AdSpR..33..638H,2007Ap&SS.308..181H,2007Ap&SS.308..191V} for recent  reviews).

After the first optical detection of the brightest one, RX J1856.4-3754 \citetext{Walter, Wolk, \& Neuh{\"a}user 1996},  \linebreak\citet{1997A&A...326..662H} found the second brightest
INS \linebreak RX\,J0720.4-3125. It was detected in the optical B and R band by \citet{1998ApJ...507L..49K}. Using deep observations Motch, Zavlin, \& Haberl (2003) derived the pro\-per motion of \rxj and measured the B magnitude with   VLT/FORS 1. Later 
{ Kaplan, van Kerkwijk, \& Anderson (2007) 
measured}  
 the distance of \rxj ($\sim 360$\,pc) and updated the pro\-per motion 
{ using} 
Hubble Space Telescope (HST) observations. 

Comparing blackbody fits derived from X-Ray observations with the flux measured in the optical/UV wavelength \citep{2003A&A...408..323M}
{ shows}  
that \rxj exceeds its expected optical/UV flux by about { an order of}  magnitude \citep{2009A&A...497L...9H}. Furthermore \citet{2006A&A...451L..17H} detected variations in the X-ray flux. The origin of these variations is still 
{ controversial} 
 \citetext{Hohle et al. 2009; Kaplan et al. 2007}.

In this paper we 
{ measure}  
the V magnitude for the first time and update the proper motion and absolute position.

In Section \ref{obs} we 
{ discuss}  
the 
{ observations}  
and data reduction. Sections \ref{phot} and \ref{relphot} addresses the photometry.  In Section \ref{sed} we discuss how our newly derived V magnitude fits with the investigations of earlier work. Section \ref{pm} is dedicated to our astrometric investigations and in Section \ref{sum} we summarize our results.

\section{Observations and Data reduction}\label{obs}
In January of 2008 we observed \rxj for about three hours exposure time at ESO Paranal observatory with the FOcal Reducer/low dispersion Spectrograph (FORS 1, \citealp{1998Msngr..94....1A}), installed in Unit Telescope Kue\-yen of the Very Large Telescope (VLT). Observations were carried out in the V Bessel filter. We used a pixel scale of $0.125''$/pixel which leads to a field of view of $4.2'\times 4.2'$. 13 single exposures of 900\,s were taken over 1.5 nights. 

\begin{figure*}
\centering
\resizebox{0.8\hsize}{!}{\includegraphics{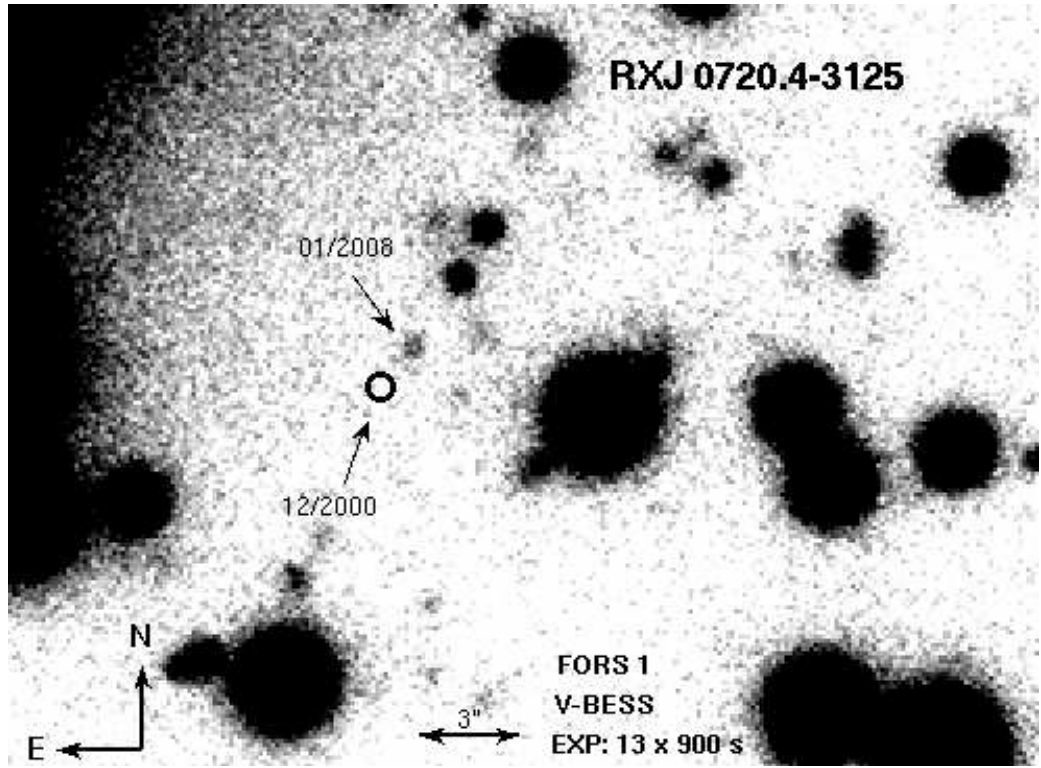}}
\caption{{ 
Co-added}  
V band images obtained with FORS\,1 on Kueyen in January 2008. The circle shows the position of \rxj as measured in the FORS\,1 image of December 2000 \citep{2003A&A...408..323M}. The arrow labeled 01/2008 points to our detected position of \rxj. 
}
\label{figure:1}
\end{figure*}

For 
{ astrometric calibration} 
we observed the globular cluster 47-Tuc with 
{ a}  
pointing at \nolinebreak{$\alpha=00^h$22$^m$29.3$^s$} and \linebreak\nolinebreak{$\delta=-71^d$59$'$54.3$''$}. The observations are summarized in Table \ref{table:1}. The 
{ image of the}  
\rxj 
{ field}  
is shown in Fig. \ref{figure:1}.

\begin{table*}
\caption{Observation log} 
\label{table:1} 
\centering 
\begin{tabular}{l l c c c c l} 
\hline\hline 
Date & Instrument & Filter & Exp & Pixel & Seeing & Target\\ 
     & telescope  &      & s  & $('')$& $('')$ \\
\hline 
{12 Jan. 2008}    & FORS1/VLT-UT1 & V Bessel & {$ 8.1\cdot 10^{3}$} & 0.125 & 1.1 & \rxj\\ 
{13 Jan. 2008}    &{ FORS1/VLT-UT1 }&{ V Bessel} &{ $3.6\cdot 10^{3}$} &{ 0.125} &{ 0.8} &{ \rxj}\\ 
12 Jan. 2008 00:42 (UT)& FORS1/VLT-UT1 & V Bessel & $4\cdot 10$   & 0.125 & 1.6 & 47-Tuc\\
12 Jan. 2008 00:47 (UT)& FORS1/VLT-UT1 & V Bessel &  12                 & 0.125 & 1.6 & Landolt field SA 98\\
12 Jan. 2008 04:48 (UT)& FORS1/VLT-UT1 & V Bessel &   5                  & 0.125 & 1.1 & Landolt field SA 98\\
\hline 
\end{tabular}
\end{table*}

\begin{figure}
\centering
\resizebox{0.99\hsize}{!}{\includegraphics[angle=270]{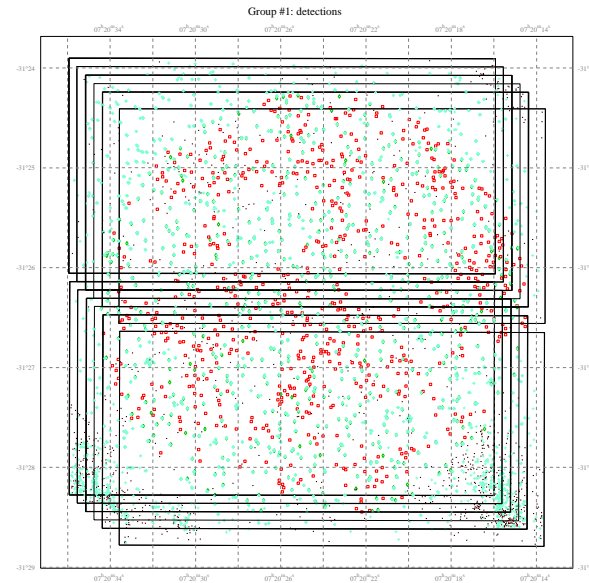}}
\caption{Illustration how SCAMP co-added the 13 FORS 1 images (two parts of the chip). { We used a dither pattern of four positions to avoid bad pixels.} Diamonds and pluses symbolize objects found in each, or at least some of the images and used for calculation of the astrometric solution while squares symbolize objects not used for calibration. The rectangles illustrate the chip shape and the relative position.}
\label{figure:3}
\end{figure}

All images were flat fielded and bias subtracted. Since FORS\,1 got a new detector in 2007 we were restricted to absolute world coordinates for astrometric comparison to earlier observations. { Furthermore the new chip is divided in two parts which limits the number of reference stars on each part. In order to derive an accurate astrometric solution we observed the globular cluster 47-Tuc at four dither positions. Based on the ESO fits header we derived image, as well as world coordinates of each object in the images with the \textit{Source Extractor} (\textit{SE}, \citealt{1996A&AS..117..393B}). These object catalogs are fitted with reference to the the two micron all sky survey catalog (2MASS, \citealt{2003tmc..book.....C}), provided by the ViZiR  database \citetext{Ochsenbein, Bauer, \& Marcout 2000}. After the WCS frame is calibrated in that way a fifth order polynomial is fitted to the data. This $\chi^2$-algorithm corrects the remaining field distortions. We used the software \textit{SCAMP}\footnote{Software for Calibrating AstroMetry and Photometry} \citep{2006ASPC..351..112B} for these calculations. \textit{SCAMP} stores the calibration information (WCS  transformation as well as field distortion coefficients) in an external header for each image. We extracted the field distortion coefficients, we derived calibrating the 47-Tuc images and used them as input for the calibration procedure of the \rxj images, which is done by \textit{SCAMP} as well in the same way. Similar, but directly without using a calibration cluster the \rxj images of the year 2000 observed by  \citet{2003A&A...408..323M} were calibrated. The field distorion map of this calibration is shown as an example in Fig. \ref{figure:2}.


\begin{figure}
\centering
\resizebox{0.99\hsize}{!}{\includegraphics[angle=270]{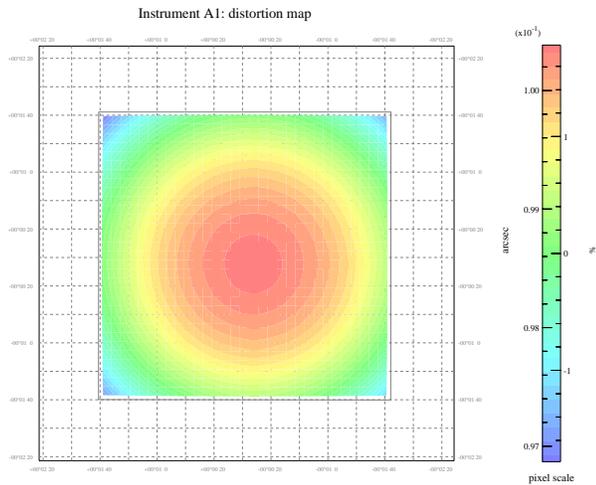}}
\caption{Field distortion map of FORS\,1 detector in December 2000. 
Color coding goes from red (middle in the image and top of the scale) through yellow and green to blue (corners of the image and bottom of the scale.)}
\label{figure:2}
\end{figure}

Another software called \textit{SWarp} \citep{2002ASPC..281..228B} uses the external headers produced by \textit{SCAMP} to align, resample, and co-add the images. A mesh based background subtraction is applied during co-addition and the two parts of the chip are stitched together. This is illustrated in Fig. \ref{figure:3}.}

{
 Typically the semi major axis of the error ellipse  of  \linebreak 2MASS does not exceed 0.3\,arcsec. The 
statistical  
errors of the source detection procedure are much smaller. \textit{SCAMP} preselects unambiguous, well exposed sources for the fitting and treats the uncertainties in a $\chi^2$ sense. Nevertheless it is 
unnecessary  
to treat each error source individually. The resulting measurement plus  calibration error is calculated in a statistical way as described in Section \ref{pm}.}

\section{Photometry for January 12th, 2008}\label{phot}
\begin{figure}
\centering
\resizebox{0.8\hsize}{!}{\includegraphics{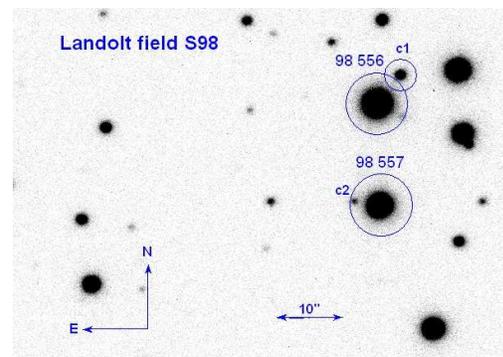}}
\caption{VLT/FORS1 image of the Landolt standard stars 98\,556 and 98\,557. There are two objects close to the standard stars, marked as c1 and c2, which contribute flux to the standard magnitude measured by \citet{1992AJ....104..340L} so we measured the flux of these objects as well and took their contribution into account.}
\label{figure:6}
\end{figure}

To 
{ flux-calibrate the image}  
we observed the Landolt standard star field SA98 \citep{1992AJ....104..340L} in the first night. { Since the standards were taken only in the first night we used the nine images of \rxj taken in the first night for standard star photometry. We co-added them in the same way as described above, but only the upper part of the chip was used this time to avoid possible systematic photometry errors as they may happen while equalizing the background of the two parts of the FORS\,1 detector.}  In the upper half of the chip only 
{ the}  
two standard stars
98\,556 and 98\,557
were 
observed at airmasses $1.656$ and $1.112$, { respectively}, 
 (Fig. \ref{figure:6}). During data analysis it turned out, that there are faint objects nearby. Using the Landolt magnitudes for the stars and the given errors we first calculated the difference between the two magnitudes of the stars as measured by Landolt and by us 
{ assuming}  
that
\[
V_{556,\,\mathrm{Land}}-V_{557,\,\mathrm{Land}}\approx V_{556,\,\mathrm{inst}}-V_{557,\,\mathrm{inst}},
\]
where $V_{\mathrm{inst}}$ are our instrumental magnitudes. 
Our measurements 
{ show}   
that the stars marked as c1 and c2 in Fig. \ref{figure:6} 
{ were}  
included in the aperture of \citet{1992AJ....104..340L} so we took them into account accordingly. Aperture photometry is done using the DAOPHOT package \citep{1987PASP...99..191S}.

We measured the instrumental magnitudes of both standard stars in each field and calculated the zero point correction $c$ and the first order extinction coefficient $k$. 
Taking into account all sources of error, which is the $3\sigma$ error of the instrumental magnitude, the error of the Landolt magnitudes and the difference between the values by using the different standard stars we derive 
\begin{eqnarray*}
c&=&(-21.0832\pm 0.0017)\,\mathrm{mag}\ \mathrm{and}\\
k&=&(0.1562\pm0.0058)\,\mathrm{mag}.
\end{eqnarray*}
{ Since these quantities are derived using two different stars, their uncertainties include an estimation of the error, which is made neglecting the color
term.}  

\begin{figure}
\resizebox{\hsize}{!}{\includegraphics{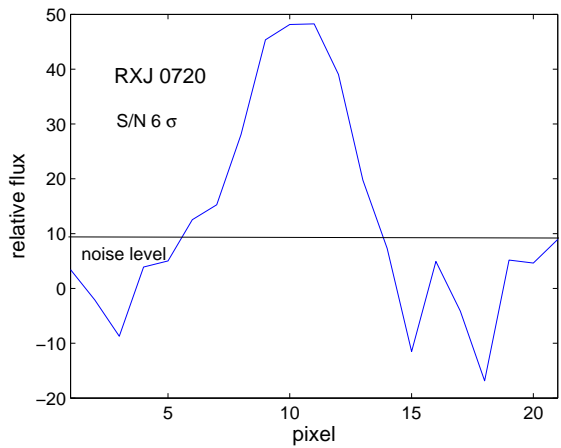}}
\caption{Projection of the detection of \rxj. The neutron star is detected with $\sim 6\sigma$.}
\label{figure:7}
\end{figure}

The most pronounced error source comes from the { faint} INS magnitude measurement itself. 
Fig. \ref{figure:1} shows the rim of an association of bright O-stars to the left of \rxj. In our deep exposures these stars influence the background, so we had to choose an aperture smaller than that in the standard star image to deal with this problem. 
The INS is detected 
{ at}  
$\sim~6~\sigma$. Because of the faintness of the source only the top of the PSF has a higher count rate than the background hence, for the INS a smaller aperture than for the reference stars is possible (Fig. \ref{figure:7}). We investigated the magnitude - aperture dependence of this faint source and concluded from our measurements and Fig. \ref{figure:7} that an aperture radius of seven pixels is sufficient. 

{ Using the uncertainties of $c$ and $k$ and taking the variation of the airmass ($Y_{\mathrm{NS}}=1.028\pm0.019$) during observation into account} we finally derive the V magnitude of \rxj to be 
\[V=(26.88\pm0.15)\,\mathrm{mag,\ for\ Jan\ 12th,\ 2008}.\]

\section{Photometry for January 13th, 2008}\label{relphot}
\begin{figure}
\resizebox{\hsize}{!}{\includegraphics{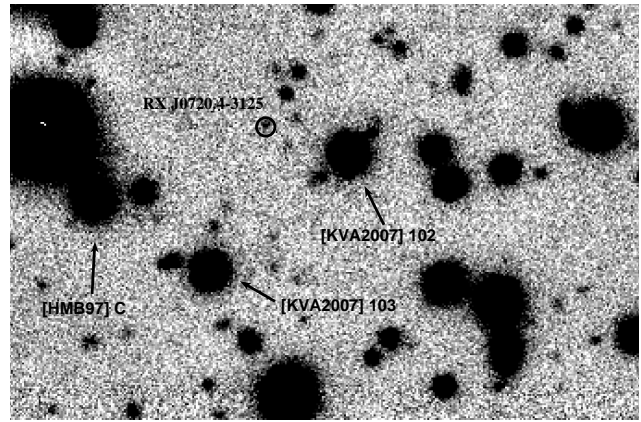}}
\caption{Co-added and background subtracted image used for relative photometry. The reference stars used are named and marked by arrows.  \cite{1998A&A...333L..59M} measured their V magnitude with an accuracy of 0.05 mag. The INS is marked and the aperture radius (5 pixel, circle) is shown.}
\label{figure:8}
\end{figure}

Four images were taken in the second night under much better seeing conditions than the first night. Since the most pronounced source of error in the absolute photometry was the measurement of the faint INSs magnitude itself, we decided to use these images { too, even though no official photometric standards were taken that night}. We co-added the images as before and explained in Section \ref{obs}. { In addition} we were using the \textit{SE} to apply a background subtraction. The \textit{SE} uses a mesh based background algorithm with a mesh size of 64 pixels. In the resulting image we identified three reference stars near \rxj with { known} V-band magnitudes (\citealt{1998A&A...333L..59M} and Fig. \ref{figure:8}).

We used an aperture 
{ of 15 pixel radius} 
for the reference stars, given by the { full width at half maximum} (FWHM) size of $\approx 4.5\,${ pixel}. For the INS we determined the 
{ object's} 
size. It turned out that the PSF of the INS vanishes below the background at a radius of less than five pixels, so we chose the aperture radius accordingly. Furthermore we chose smaller inner and outer sky radii compared to the reference stars to 
{ prevent} 
the neighboring stars 
{ from falling into the background annulus. } 

The difference of the instrumental and the apparent \linebreak magnitudes for each reference star is calculated. The mean of this difference gives again the detector zero point, { so} this the magnitude of \rxj is determined to be
 \[
 V=(26.81\pm 0.09)\mathrm{mag,\ for \ Jan\ 13th,\ 2008}
 \]
 which is fully consistent with our other measurement.
 
Note, that if the small aperture would have caused any 
loss of flux  
the magnitude of the INS would be { larger}  instead. Since 
the error bar is smaller we use the relative value in the following discussion.  

\section{The Spectral Energy Distribution of \rxj}\label{sed}
The visual magnitude of \rxj 
{ is consistent  with} 
an optical excess 
{ about an order of} magnitude 
{ larger than}  
the expected flux 
{ extrapolated} 
from the X-ray spectra first reported in \citet{1998ApJ...507L..49K} and studied in detail by \citet{2003ApJ...590.1008K}. However, the origin of this enigmatic property as well as its relation to the X-ray variations is still unknown and a matter of debate, see \citet{2007Ap&SS.308..181H} and \citet{2006A&A...451L..17H} for a review about the INSs and \rxj in particular. It might be possible that the emitting area of the soft X-ray radiation is not strictly connected to the source of optical/UV photons \citep{2009A&A...497L...9H}. { \citet{2005esns.conf..117T}; Turolla, Zane \& Drake (2004); Zane, Turolla \& Page (2007); Zane, Turolla \& Drake (2004) and references therein discuss alternative explanations of the optical excess.} The radius of the X-Ray emitting area is $\approx4.5$\,km \citep{2006A&A...451L..17H}, normalized to a distance of 300pc. 

\begin{table*}
\caption{Optical, ground based observations of RX\,J0720.4-3125. Only measurements of ground based observations are shown, in contrast to Fig. \ref{plot}, where HST observations from \citet{2003ApJ...590.1008K} are included. Other references are given in the table.} 
\label{table:3} 
\centering 
\renewcommand{\arraystretch}{1.5}
\begin{tabular}{lr@{}l@{\,$\pm\,$}rr@{}r@{.}lrr@{\,}r@{.}ll} 
\hline\hline 
Filter & \multicolumn{3}{c}{Mag} & \multicolumn{3}{c}{Flux density} & $\lambda_{\mathrm{cent}}$ & \multicolumn{3}{c}{Flux} & Refernence\\
& \multicolumn{3}{c}{\,}&\multicolumn{3}{c}{[W/m$^2$/Hz] $\cdot 10^{-34}$}&[\AA] & \multicolumn{3}{c}{[erg s$^{-1}$ cm$^{-2}$] $\cdot 10^{16}$}&\\
\hline                                             	       
U	&& 25.68  &0.17    &&   9&75$^{+1.65}_{-1.42}$   & 3600 && 8&$13^{+1.38}_{-1.18}$           & Motch et al. 2003\\
B	&& 26.58  &0.25    &&   9&64$^{+2.46}_{-1.98}$   & 4300 && 6&$73^{+1.72}_{-1.38}$          & Motch et al. 2003\\
 	&& 26.79  &0.20    &&   7&94$^{+1.61}_{-1.33}$   & 4300	 && 5&$54^{+1.12}_{-0.93}$       & Motch et al. 2003\\
 	&& 26.44  &0.15    && 11&00$^{+1.60}_{-1.45}$   & 4300	 && 7&$67^{+1.12}_{-1.01}$         & Motch et al. 2003\\
 	&& 26.787&0.040  &&   7&96$^{+0.30}_{-0.28}$   & 4300	 && 5&$55^{+0.21}_{-0.20}$     & Motch et al. 2003\\
 	&& 26.620&0.050  &&   9&29$^{+0.44}_{-0.42}$   & 4300	 && 6&$48^{+0.31}_{-0.29}$     & Motch et al. 2003\\
V	&& 26.81  &0.09    &&   6&69$^{+0.99}_{-0.86}$   & 5500 && 3&$65^{+0.54}_{-0.47}$     & this paper\\
R     &&  26.9    &0.3      &&   5&11$^{+1.63}_{-1.23}$   & 7000 && 2&$19^{+0.70}_{-0.53}$        & Kulkarni \& van Kerkwijk 1998\\
H&$>$& 23.07&0.11&$<$&   7&75$^{+0.80}_{-0.72}$   &16500&$<$&12&$29^{+1.32}_{-1.12}$ & Posselt et al. 2009 (upper limit)\\
\hline
\end{tabular}
\end{table*}

\begin{figure}
\centering
\resizebox{0.93\hsize}{!}{\includegraphics{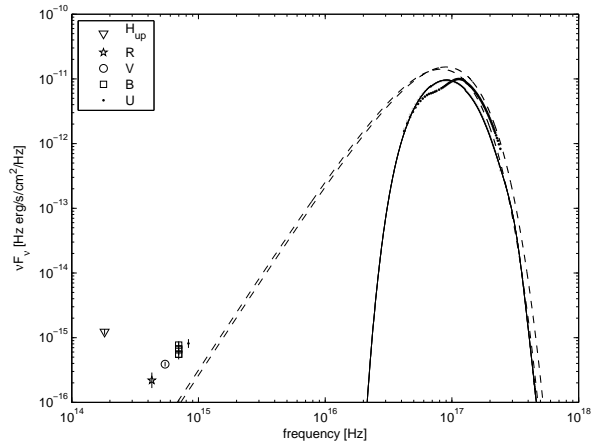}}
\caption{Optical/UV flux from R to UV magnitudes (our new V-band data point is marked as circle) compared to X-Ray spectra obtained from \xmm\ EPIC-pn (right) performed in full frame mode with thin filter. The two spectra of \rxj\ are taken from orbit 0078 {(grey dots)} and 0815 {(black dots)} with effective temperatures of 86.5$\pm$0.4eV and 94.6$\pm$0.5eV, respectively. { The solid line marks the best fit model for rev. 0078 (including interstellar
extinction), while the dashed lines corresponding to the black bodies for both
spectra without any absorption.} The broad absorption feature near 300eV ($\widehat{=}7.25\cdot 10^{16}$\,Hz) can be seen { in the spectrum of revolution 0815 (black dots)}. The Optical magnitudes are sumarized in table \ref{table:3}, { the flux in the H band} is an upper limit.}
\label{plot}%
\end{figure}

For comparison to other optical/UV magnitudes we \linebreak show in Fig. \ref{plot} the flux expected from a black body model (\xmm\ EPIC-pn). The effective temperature of \linebreak \rxj is changing 
{ on} 
time scales of { years}, { see \citet{2009A&A...498..811H} and \citet{2006A&A...451L..17H}.}

We plot in Fig. \ref{plot} the spectra obtained { with \linebreak \xmm\ EPIC-pn} with lowest (orbit 0078) and  highest (orbit 0815) temperatures ($86.5\pm0.4$\,eV and  $94.6\pm0.5$\,eV, respectively) measured yet. The values were  ob\-tain\-ed by fitting the spectra with standard software \textit{xspec12} \linebreak\citetext{Dorman, Arnaud, \& Gordon 2003} in the energy range of 0.16-1.5\,keV by applying a black body model with addi\-tive gaussian absorption line ($\mathrm{line\ energy}=301\pm3$\,eV, $\sigma=77\pm2$\,eV) with absorption due to the ISM ($N_{H}=1.04\pm0.02\cdot10^{20}$\,cm$^{-2}$), see \citet{2009A&A...498..811H} for details. 
  
\section{Astrometry}\label{pm}
\begin{figure*}
\centering
\resizebox{0.93\hsize}{!}{\includegraphics{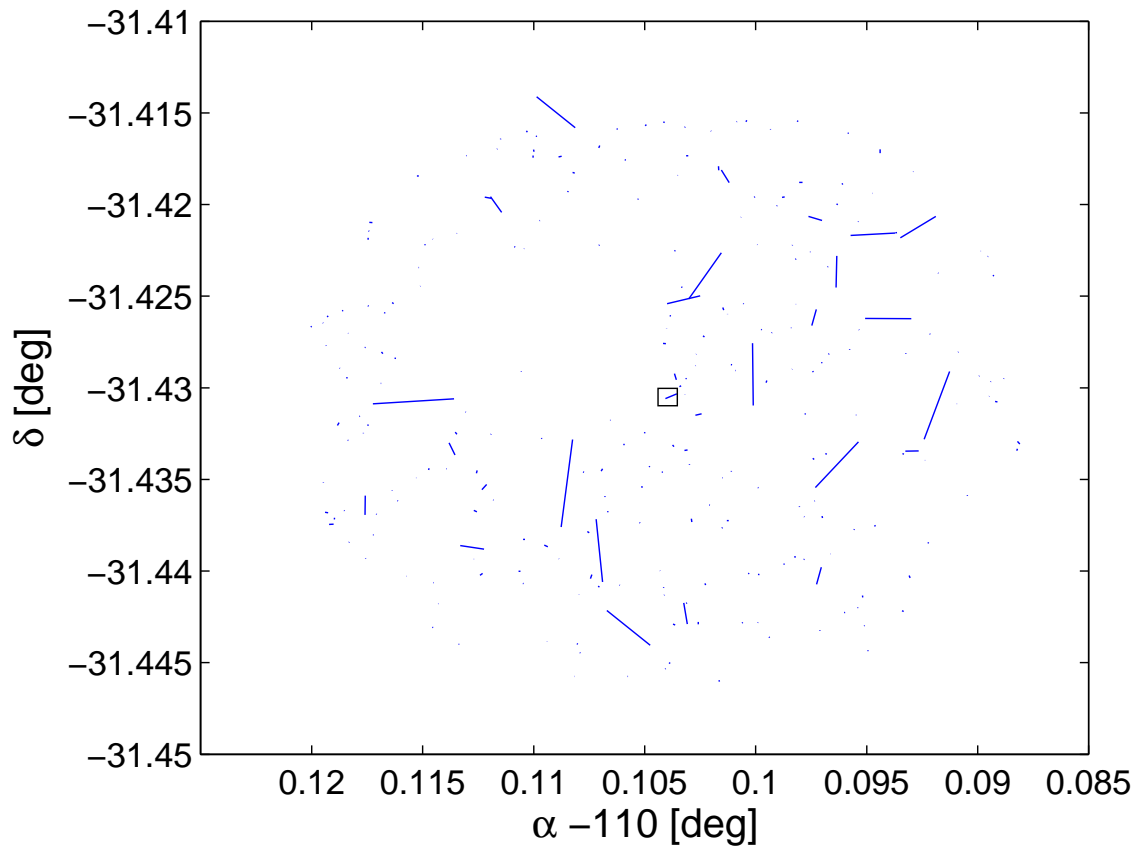}}
\caption{Detected objects within a radius of 1' around the position of \rxj. The lines indicate the change of position between the two images of { MJD 51902} and { MJD 54477} of the objects, scaled up by a factor of 20. While \rxj (middle) is obviously moving, most of the other objects are almost standing still, demonstrating the high accuracy of the calibration. Other ''fast moving'' objects are false detections or wrong assignments of objects between the two images.}
\label{figure:4}
\end{figure*}

For the longest possible epoch difference we used the B-band image of \citet{2003A&A...408..323M} from the year 2000 and compared the position of all objects with the positions in our image from 2008. The objects in the two images are assigned to each other and the change of position is calculated (Fig. \ref{figure:4}). 

Afterwards the change of position undergoes a Kol\-mo\-go\-rov-Smir\-nov 2-sample test comparing the absolute position changes of the objects
to a simulated Rayleigh distribution. The test is performed comparing the cumulative distribution functions (CDF) of the test sample and the data. 
If the test fails objects lying outside 2$\sigma$ from the mean of the distribution are excluded. The mean and standard deviation of the distribution are recalculated and the motion of all objects is shifted by that mean to reduce the last remaining systematical errors. This is repeated until the test succeeds. What is left is a sample of Rayleigh distributed background stars, without any 
{ systematic}  
 effects. The standard deviation $\sigma_{\mathrm{Back}}$ of these stars could be seen as total calibration error. Since the NS is much fainter than the background stars, the positional error of its photo-center is 
{ relatively large. We} 
added the intrinsic proper motion error $\Delta\mu_{\mathrm{intr}}$ (which is again calculated by the position error in both epochs and the epoch difference) of the NS in both epochs to that error, representing the full error of the neutron stars motion as
\begin{displaymath}
\Delta \mu_{\mathrm{NS}}=\sqrt{(\Delta\mu_{\mathrm{intr}})^2+(\sigma_{\mathrm{back}})^2}.
\end{displaymath}
The mean intrinsic error is already included in $\sigma_{\mathrm{Back}}$ but is larger for the barely visible NS than for the background objects.

{ We}  
derive a proper motion value 
\begin{eqnarray*}
\nonumber
\mu_{\alpha}&=& (-93.2\pm5.4)\,\mmas/\mathrm{yr}\\\nonumber
\mu_{\delta}&=& (48.6\pm5.1) \,\mmas/\mathrm{yr}\\
\mu &=& (105.1\pm7.4) \,\mmas/\mathrm{yr}.\\\nonumber
\theta&=&(296.951\pm0.0063)^{\circ}\nonumber
\end{eqnarray*}
This proper motion is consistent with previously published values \citep{2003A&A...408..323M,2007ApJ...660.1428K}.
\linebreak \rxj changed its position from December 2000 ({ $\mathrm{MJD}=51902$\,days}) to January 2008 ({ $\mathrm{MJD}=54477$ days}) from
\begin{displaymath}
\begin{array}{r@{\,}l@{\,}ll@{\,}l@{\,}ll}
\mathrm{MJD}&\ =&51902\mathrm{\,days:}&&&&\\
7^h&20^m&24.976^s&-31^d&25'&50.105''&\mathrm{to}\\
   & \pm   &14.15\,  \mmas&& \pm  &12.82\,  \mmas&\\
\mathrm{MJD}&\ =&54477\mathrm{\,days:}&&&&\\
7^h&20^m&24.926^s&-31^d&25'&49.776''&\\
   &   \pm &19.78\,  \mmas&&  \pm &19.30\,  \mmas,&
\end{array}
\end{displaymath}
Since the { total exposure} time in the year 2000 image was longer and the observation wavelength was shorter, our \linebreak new\-ly reduced position measurement of the year 2000 is more precise
{ and} 
the signal to noise ratio is better.

\begin{figure}
\resizebox{\hsize}{!}{\includegraphics{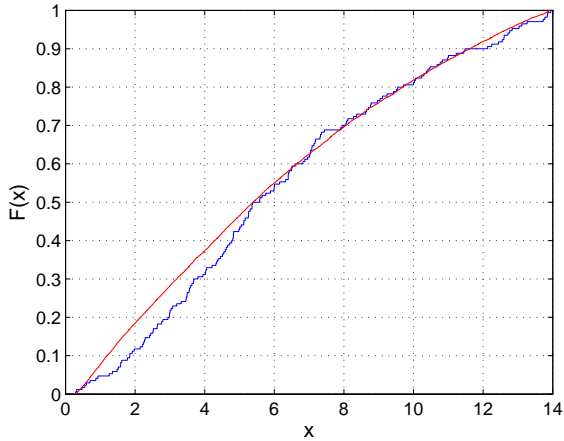}}
\caption{Cumulative distribution function (CDF) of the Rayleigh distributed background stars in the \rxj field, compared with the CDF of a synthetic Rayleigh distribution. The comparison shows good agreement, proving that there are no systematical errors left in the calculation.}
\label{figure:5}
\end{figure}

\section{Summary}\label{sum}
Taking a deep optical image of \rxj we were able to determine the V magnitude and updated the position.
{ Our 
proper motion ($\mu=105.1\pm7.4$\,mas) confirms the measurement obtained by \citet{2007ApJ...660.1428K} with the HST ($107.8\pm1.2$\,mas)} with a difference of less than 3 mas/yr.

Our V magnitude { 
($\mathrm{V}=26.81\pm0.09$\,mag) is compatible with previous observations from \citet{2003A&A...408..323M} ($\mathrm{B}=26.620\pm0.050$\,mag) and from \citetext{Kulkarni \& van Kerkwijk (1998)} ($\mathrm{R}=26.9\pm0.3$\,mag). These data can be fitted by the emergent spectrum of the photoionized plasma, surrounding the NS \citep{2009A&A...497L...9H}. This scenario would also explain the power-law component of the "optical excess" \citep{2003ApJ...590.1008K}, which is confirmed by our data.} 

For further investigations correlated observations in X-ray and the optical/UV are needed to collect enough data to investigate the nature of the flux variations of \rxj and to understand this isolated neutron star. 

We also note, that there are no additional faint, fast moving objects detected in this field (by comparing the year 2000  observations and our image), hence no additional similar NS in this field.

\acknowledgements{The authors would like to thank Frank \linebreak Haberl, Fred Walter, Andreas Seifahrt and Tristan R\"oll for fruitful discussions and improving comments.
Furthermore we thank the Staff members of ESO Paranal observatory for their help and assistance carrying out the observations.
Based on observations made with ESO Telescopes at the La Silla or Paranal Observatories under programme ID 66.D-0286(A). 
Moreover we thank Emmanuel Bertin and his team for providing new data reduction software. This research has made use of the VizieR catalogue access tool, CDS, Strasbourg, France. This work was supported in part
by DFG grant SFB/Transregio 7 ''Gravitational Wave Astronomy''. CG acknowledges support by DFG grand under project NE 515/30-1. The work of MMH has been supported by CompStar, a research networking programme
of the European Science Foundation (ESF). TOBS acknowledges support from Evangelisches Studienwerk e.V. Villigst.

\end{document}